%Paper: cond-mat/9307013
%From: donatis@tsmi19.sissa.it
%Date: Thu, 08 Jul 1993 12:56:50 +0100
%Date (revised): Fri, 09 Jul 1993 12:04:09 +0100

\magnification=1200
\baselineskip16pt

\font \titlefont=cmr10 scaled\magstep1

\def\vf{\vfill\eject}

\null
\vskip-2truecm
\rightline{{\bf  S.I.S.S.A. 101/93/EP }}
\vskip2truecm
\vskip0.5truecm
\centerline{\titlefont  EFFECTIVE THEORY OF A CHIRAL SUPERFLUID}

\vskip3truecm
\centerline{Pietro Donatis and Roberto Iengo}
\vskip 1.0truecm
\centerline{{\it International School for Advanced Studies,
                I-34014 Trieste, Italy and}}
\centerline{{\it Istituto Nazionale di Fisica
                Nucleare, INFN, Sezione di Trieste, Trieste, Italy}}

\vskip3.0truecm
\noindent{\bf Abstract}
\vskip.3truecm
\par
\noindent
We consider an effective Lagrangian describing a fluid living on
two-di\-men\-sio\-nal planes. The fluid self-interacts through a Chern-Simons
vector potential, whose field strength is proportional to the density
fluctuation. This effective Lagrangian can be related to the Anyon
mean field, but can also be considered more generally to describe a
universality class of superfluids and, when charged, of
superconductors. We study the relevant physical properties, including
the spectrum, the chirality features appearing in the polarization of
scattered EM waves, and the peculiar response under a magnetic field,
i.e. a peculiar kind of anisotropic Meissner effect.
\vf\eject

\def\today{\ifcase\month\or
	January\or February\or March\or April\or May\or June\or
	July\or August\or September\or October\or November\or
	December\fi
	\space\number\day, \number\year}

\newcount\hour \newcount\minute
\hour=\time  \divide \hour by 60
\minute=\time
\loop  \ifnum \minute > 59 \advance \minute by -60 \repeat
\def\writetime{\ifnum \hour<13 \number\hour:%  % supresses leading 0's
                      \ifnum \minute<10 0\fi%  % so add it it
                      \number\minute
                      \ifnum \hour < 12 \ A.M. \else \ P.M. \fi
      \else \advance \hour by -12 \number\hour:%  % supresses leading 0's
                      \ifnum \minute<10 0\fi%     % add it in
                      \number\minute \ P.M. \fi}

\magnification=1200

\topskip=25pt

\font\twelvebf = cmbx12 scaled \magstep2
\font\bigbf=cmbx10 scaled \magstep2

\def\chapter#1#2{\noindent
{\bf {\twelvebf #1}\ \ {\twelvebf #2}}
\bigskip\nobreak\noindent}

\def\subchapter#1#2{\bigskip\bigskip{\noindent
{\bigbf #1}\ \ {\bigbf #2}}\bigskip\nobreak\noindent}

\def\parag#1#2{\goodbreak\medskip\medskip
                    \centerline{\bf #1\ \ #2}
                   \medskip\nobreak\noindent}

\def\appendix#1#2{\noindent
{\bf {\twelvebf Appendix #1}}\vskip15pt\noindent {\twelvebf #2}
\bigskip\bigskip\noindent}

\baselineskip=14pt
\hfuzz=20pt

\def\p{\par\noindent}
\def\v{v^2}

\def\ie{{\it i.e.}\ }
\def\vp{(v^2-|\phi|^2)}
\def\D{\vec D}
\def\A{\vec A}
\def\dps{|\D\phi|^2}
\def\II{\int d^2r\,}
\def\At{A_\theta}

\def\mr{\omega r^2}
\def\emr{e^{-\omega r^2}}
\def\Aa{\bar A}
\def\aA{\bar a}

\def\zz{\bar z}
\def\pp{\bar \partial}
\def\pa{\partial}

\def\U{\vec u}
\def\AT{\tilde A}
\def\AaT{\skew6\bar{\AT}}
\def\BT{\tilde B}
\def\x{\vec x}

\def\E{\vec E}
\def\B{\vec B}
\def\ACS{\A^{CS}}
\def\Aem{\A^{em}}
\def\aCS{A^{CS}}
\def\aem{A^{em}}

\def\gsim{ \lower .75ex \hbox{$\sim$} \llap{\raise .27ex \hbox{$>$}} }
\def\lsim{ \lower .75ex \hbox{$\sim$} \llap{\raise .27ex \hbox{$<$}} }

\def\mincir{\raise -2.truept\hbox{\rlap{\hbox{$\sim$}}\raise5.truept
\hbox{$<$}\ }}
\def\magcir{\raise -2.truept\hbox{\rlap{\hbox{$\sim$}}\raise5.truept
\hbox{$>$}\ }}

\newcount\equanumber	\equanumber=0
\def\chapterlabel{1}
\def\eqname#1{\relax\global\advance\equanumber by 1
  \xdef#1{{\rm(\chapterlabel.\number\equanumber)}}#1}

\def\eqn{\eqno\eqname}

\newcount\refnumber	\refnumber=0
\def\refname#1{\relax\global\advance\refnumber by 1
  \xdef#1{{\rm[\number\refnumber]}}#1}
\def\ref{\refname}

\chapter{1.}{Introduction and summary}

We study here the physical implications of an effective Lagrangian \`a la
Landau-Ginzburg, describing a possible peculiar superfluid. We are
interested in particular in the case when the superfluid is charged,
behaving thus as a peculiar superconductor. The distinctive features of
this superfluid are, first the fact that it lives on a two dimensional
surface, and this is in common with various theoretical studies of layered
superconductors (systems essentially made of two dimensional layers over
which the supercurrent can flow). Also, many high $T_c$ superconductors
appear to behave in this way.
\par
The other, more peculiar, feature is chirality; that is the fact that the
dynamics of this superfluid makes a distinction between some left and right
handed behaviour. This is built in the defining effective Lagrangian, which
breaks parity, although the consequences may not be immediate to see. In
fact, the effective Lagrangian contains a peculiar self-interaction
expressed in terms of a Chern-Simons gauge field. One can thus phrase our
investigation as a study of a universality class of some superfluids, which
can be related to Anyon systems. Of course, one can also forget about
Anyons, and consider the present superfluid {\it per se} as a possible physical
system.
\par
The main point about the Chern-Simons gauge field is that we assume that
its field strength (also called ``curvature'') is determined by the request
that it is proportional to the density fluctuation, \ie the actual density
at a given point minus the average density. This assumption makes the
effective Lagrangian consistent, and, in particular, compatible with
translational invariance.
\par
Formally, the same effective Lagrangian has been introduced in refs.
\ref\zhk\ \ref\leezhang\ \ref\leefisher\ \ref\zhang, to provide a possible
description of the Fractional Quantum Hall Effect (FQHE). Here, on the
contrary, we do not assume a strong, uniform external magnetic field,
orthogonal to the surface, like in the FQHE. Thus, our investigation is
rather devoted to the superfluidity and the possible superconductivity
properties. Essentially the same effective Lagrangian also been previously
considered in ref. \ref\Boya\ in a generic context of Anyon
superconductivity. Our treatment of the problem is however basically
different from what discussed in ref. \Boya.
\par
We devote an appendix (Appendix A) to the relation of our effective
Lagrangian with the translationally invariant version of the Anyon Mean
Field theory developed in refs. \ref\Kurt\ \ref\Lidingping\ \ref\Lechner\
\ref\Iengo.
\par
In Section 2 we begin by discussing the effective Lagrangian and the
spectrum of the excitations it describes, first by a perturbative treatment
and then by studying also the vortex excitations. We find a spectrum with a
finite gap, and this gap is of course the same as that found in ref.
\zhang. However there is an important difference in the range of values of
a parameter, having the meaning of a selfinteraction coupling constant.
Here we are interested in a range of values for which we show that the
``roton'' (or vortex) part of the spectrum is above the gap, contrary to
the case relevant for the FQHE considered in ref. \zhang. In fact, our
range of values of the parameter is such that when  the effective
Lagrangian is compared with Anyon Mean Field (Appendix A), all the excitations
correspond to the inter Landau levels ones, rather than the intra Landau
levels as relevant for FQHE. Thus, in our case, the spectrum resembles the
one of an ordinary superfluid, shifted up by the gap amount, as shown in
Fig. 1.
\par
We then discuss in Section 3 the chirality properties. In particular we
show that a superfluid layer has an optical activity, in that it transmits
only circularly polarized electromagnetic waves of a definite handedness,
in a range of frequencies near to the above mentioned gap. Qualitatively,
those frequencies are in the microwave/infrared regime (for reasonable
values of the parameters), where it would be measurably difficult to
detect, and we show that the effect disappears at high frequencies.
\par
In Section 4 we discuss the magnetic properties of the chiral superfluid,
discussing the possibility of the Meissner effect in the bulk of a sample
made of a stack of many layers (see Fig. 2{\it a} and 2{\it b}). We find a
very different behaviour, depending on the orientation of the magnetic
field relative to the layers. If the field is orthogonal to the layers,
case {\it a}, then the system behaves as a type II superconductor with a
finite penetration length. We show it by considering both the penetration
from an edge and also studying magnetic vortices. In order to get the
correct result one has to include in the computation the electrostatic
forces, which of course are present and important in a charged superfluid
in general, but are particularly essential in our case.
\par
In the case {\it b}, where the magnetic field is parallel to the layers, the
behaviour is very different. One has to make a further distinction between
to cases: the field, which is parallel to the layers' plane, can either try
to penetrate {\it i}) in the direction parallel to the layers or else
{\it ii}) in the direction orthogonal to the layers. In the first case {\it
i}),
the behaviour is determined, like in the layered superconductors
studied in ref. \ref\Coffey\ by the amount of the interlayer Josephson
coupling. Namely if the Josephson coupling is small the penetration length
in the interlayer space can be large. This is the same as for ordinary
superconductors.
\par
In the second case {\it ii}) the chiral superconductor behaves differently
from an ordinary one. We show, by an order of magnitude argument, that in
this case the penetration length, in the direction orthogonal to the
layers, is not finite and independent of the size of the sample. Rather, it
grows with the power 1/3 of the sample's size, apart from logarithmic
corrections.
\par
Summarizing, the effective Lagrangian we have studied
describes a possible physical system which is rich of rather interesting
structures.

\bigskip
\bigskip
\bigskip

\newcount\equanumber	\equanumber=0
\def\chapterlabel{2}
\def\eqname#1{\relax\global\advance\equanumber by 1
  \xdef#1{{\rm(\chapterlabel.\number\equanumber)}}#1}

\def\eqn{\eqno\eqname}

\chapter{2.}{Effective theory}

We take as a starting point the following non relativistic effective
Lagrangian density in two space and one time dimensions:
$$
{\cal L}=i\phi^{*}\pa_0\phi-
{1\over 2m}\bigl|\vec D\phi\bigr|^2-
g|\phi|^4=
i\phi^{*}\pa_0\phi-{\cal H}
\eqn\lagr
$$
here $\phi (\vec x,t)$ is a non relativistic complex field which plays the
r\^ole of order parameter, related to the density by
$$
\rho=|\phi|^2
\eqn\
$$
We assume a fixed total number of particles $N$, therefore we keep {\it fixed}
$N\!=\!\int d^2x\,|\phi|^2$.
\p
$\vec D$ is the covariant derivative
$\vec D\!=\!\vec\nabla\!-\!i\skew6\vec{\AT}$ and $\AT$ is the ``fluctuation''
of a CS gauge potential determined by:
$$
\vec\nabla\wedge\skew6\vec{\AT}={2\pi\over k}\delta\rho
\qquad \delta\rho=\rho-\v=|\phi|^2-\v
\qquad \v=<\rho>
\eqn\bcs
$$
As a consequence of the conservation of the number of particles we have
consistently:
$$
\int d^2x\,\vec\nabla\wedge\skew6\vec{\AT}=0
\eqn\cons
$$
One can compare equation \bcs\ with the
standard definition of the CS field describing Anyons, where the CS field
strength is proportional to the density. Our effective Lagrangian
\lagr\ is meant to describe the physics of the fluctuations over the
constant density, and therefore the dynamical effects of the CS
fluctuations as in equation \bcs.
\p
In Appendix A we show how to relate the Lagrangian \lagr\ to the Anyon Mean
Field theory.
\p
For convenience we choose to change the zero of the energy spectrum adding
to the Hamiltonian density the constant quantity $-gv^4$. So that in
\lagr\ ${\cal H}$ becomes:
$$
{\cal H}={1\over 2m}\bigl|\bigl(\vec\nabla-i\skew6\vec{\AT}\bigr)\phi\bigr|^2+
g(|\phi|^4-v^4)=
{1\over 2m}\bigl|\bigl(\vec\nabla-i\skew6\vec{\AT}\bigr)\phi\bigr|^2+
g(\delta\rho)^2
\eqn\
$$
here we used the fact that $\int d^2x\,\delta\rho\!=\!0$ in the Hamiltonian.
\goodbreak
\subchapter{2.1.}{Small deformations approach}
\noindent
The first analysis we want to perform on Lagrangian \lagr\ is based on a small
deformation approach. Let us take the following parameterization:
$$
\phi=fe^{i\theta}
\eqn\param
$$
Since the integral over all the plane of $\delta\rho (\x,t)$ is zero,
we can write it as the divergence of some quantity, and we choose:
$$
\delta\rho(x)=2v\vec\nabla\cdot\U (\x,t)
\eqn\fluc
$$
furthermore we have the gauge freedom to take $\U (\x,t)$ irrotational, that is
$$
\vec\nabla\wedge\U=\pa_x u_y-\pa_y u_x=0
\eqn\irrot
$$
Then we have from \bcs\ and \fluc:
$$
\vec\nabla\wedge\skew6\vec{\AT}={4\pi v\over k}\vec\nabla\cdot\U
\eqn\
$$
this equation can be solved by:
$$
\AT_x=-{4\pi v\over k}u_y \qquad \AT_y={4\pi v\over k}u_x
\eqn\afluc
$$
Notice that $\vec\nabla\cdot\skew6\vec{\AT}\!=\!0$.
\p
With the new parameterization the Lagrangian becomes (neglecting higher
orders):
$$
{\cal L}=
2v\theta\pa_0(\vec\nabla\cdot\U)-
{1\over 2m}\v(\vec\nabla\theta)^2-
{1\over 2m}\biggl({4\pi\v\over k}\biggr)^2\U^2-
{1\over 2m}(\triangle\U)^2-
4g\v(\vec\nabla\cdot\U)^2
\eqn\nlagr
$$
Performing the variation with respect to $\theta$ we get:
$$
2v\pa_0(\vec\nabla\cdot\U)+{\v\over m}\vec\nabla\cdot\vec\nabla\theta=0
\qquad \Rightarrow \qquad
{\v\over m}\vec\nabla\theta+2v\dot{\U}=0
\eqn\
$$
which can be regarded as the continuity equation:
$$
\pa_0\rho+\vec\nabla\cdot(\rho\vec V)=0
\eqn\
$$
provided we identify the velocity $\vec V\!=\!{\vec\nabla\theta\over m}$.
\p
Inserting back this equation in the Lagrangian \nlagr\ we get:
$$
{\cal L}=2m{\dot{\U}}^2-{1\over 2m}\biggl({4\pi\v\over k}\biggr)^2\U^2-
{1\over 2m}(\triangle\U)^2-
8g\v(\vec\nabla\cdot\U)^2
\eqn\smdeflag
$$
so we have managed to write the Lagrangian as a kinetic part minus a
potential part. Taking $\U\!=\!\U_0e^{i(Et+\vec p\cdot\vec x)}$ we find
the spectrum of the energy:
$$
2mE^2={1\over 2m}\biggl({4\pi\v\over k}\biggr)^2+{1\over 2m}(p_x^2+p_y^2)^2+
8g\v(p_x^2+p_y^2)
\eqn\spec
$$
Here we are neglecting a contribution to the energy coming from an
electrostatic interaction between fluctuations, which will play an
essential r\^ole in the next chapter. The piece to be added to the
Hamiltonian density is
$$
{e^2\over 8\pi}\delta\rho(\x)\int d^2x'{1\over |\x-\x'|}\delta\rho(\x');
$$
its contribution to the r.h.s. of equation \spec\ is
$+e^2\v|\vec p|$.
\p
The minimum of the energy, for $p\!=\!0$, is not zero, so we have a gap:
$$
E(0)={2\pi\over mk}\v\equiv{\cal E}
\eqn\gap
$$
Also currents can be computed:
$$
\eqalign{J_x&={1\over 2mi}\Bigl[\phi^{\dag}D_x\phi-\phi(D_x\phi)^{\dag}\Bigr]
={\v\over m}\Bigl(\pa_x\theta+{4\pi\over k}vu_y\Bigr)\cr
J_y&={1\over 2mi}\Bigl[\phi^{\dag}D_y\phi-\phi(D_y\phi)^{\dag}\Bigr]
={\v\over m}\Bigl(\pa_y\theta-{4\pi\over k}vu_x\Bigr)\cr}
\eqn\adriano
$$
which, in a compact form, can be rewritten as:
$$
J_i=-2v(\dot u_i-{\cal E}\epsilon_{ij}u_j)
\eqn\
$$
This current exhibits the {\it chiral property} of our system. The chiral
effects appear for $p\!\to\!0$.
\p
To see this let us parameterize $\U(\x,t)$ as follows:
$$
u_x(\x,t)=u_{0x}\cos(Et+\varphi_x) \qquad u_y(\x,t)=u_{0y}\cos(Et+\varphi_y)
\eqn\
$$
then if $E\!=\!{\cal E}$ \ie $p\!=\!0$, the currents \adriano\ can be
rewritten:
$$
J_x=J_0\cos({\cal E}t+\varphi_0) \qquad J_y=J_0\sin({\cal E}t+\varphi_0)
\eqn\
$$
where $J_0$ and $\varphi_0$ are constants depending on $\U_0$.
\p
Notice that this is a circularly polarized current with a definite
direction of rotation. It can be easily seen that, since such direction
depends on the sign of $k$, it reverses performing a parity transformation.
In our case $k$ is fixed since the Lagrangian \lagr\ describe a physical
system which is not parity invariant.

\subchapter{2.2.}{Vortex excitations}
\noindent
Now we go beyond the small deformations and study the vortex solutions,
\ie vortex configurations, that minimize the energy:
$$
E=\II \biggl\{ {1\over 2m} \dps+g\vp^2\biggl\}
\eqn\ren
$$
We will follow the method of reference \ref\paul\ \ref\hong\ \ref\jack\ based
on the classical work of Bogomol'nyi \ref\bog. Vortices are classical static
solutions of the form:
$$
\phi(r,\theta)=f(r)e^{in\theta} \qquad \lim_{r\to\infty}f(r)=v
\eqn\vortex
$$
here $n$ is a topological invariant corresponding to the {\it vorticity},
that is how many times the vortex winds round; the elementary vortex has
$|n|\!=\!1$.
\p
Now since in polar coordinates we have (we call here simply $A$ what we
called $\AT$ in \lagr):
$$
B={1\over r}\pa_r (r\At)-\pa_{\theta} A_r
\eqn\
$$
we can choose for the gauge field the form, see equation \bcs:
$$
\left\{
\eqalign{&A_r=0\cr
&\At=
-{2\pi\over k}{1\over r}\int_0^r dr'r'\Bigl(\v-f^2(r')\Bigr)\cr}\right.
\eqn\
$$
\p
With this choice the total energy becomes:
$$
E=\II \biggl\{ {1\over 2m}\Bigl[(\pa_r f)^2+
{1\over r^2}(n-r\At)^2 f^2\Bigr]+g\vp^2\biggl\}
\eqn\toten
$$
from this equation we get the following requirement for the energy to be
finite:
$$
\lim_{r\to\infty}\At(r)={n\over r}\rightarrow 0
\eqn\
$$
For our particular case this yields:
$$
{2\pi\over k}\int_0^{\infty} dr\,r\Bigl[\v-f^2(r)\Bigr]=-n
\eqn\
$$
which is nothing but the quantization of the magnetic flux in integer
factors of $2\pi$.
\p
The problem now is to minimize the total energy; for this purpose it is
useful the following identity \ref\jackpi:
$$
\dps =|(D_x\pm iD_y)\phi|^2\pm m\vec\nabla\wedge\vec J\pm B|\phi|^2
\eqn\
$$
here $\vec J$ is the usual current
$$
\vec J ={1\over 2mi}\Bigl[\phi^{\dag}\D\phi-\phi(\D\phi)^{\dag}\Bigr]
\eqn\current
$$
\p
Then we have:
$$
E=\II \biggl\{{1\over 2m}|(D_x\pm iD_y)\phi|^2\pm
{1\over 2}\vec\nabla\wedge\vec J\pm
{1\over 2m}B|\phi|^2+g\vp^2\biggr\}
\eqn\
$$
Due to \bcs, we can rewrite the third term as:
$$
{1\over 2m}B|\phi|^2={\pi\over mk}\vp^2+{\v\over 2m}\vec \nabla\wedge\A
\eqn\
$$
so we are left with:
$$
E=\II \biggl\{{1\over 2m}|(D_x\pm iD_y)\phi|^2\pm
{1\over 2}\vec\nabla\wedge\biggl(\vec J\pm {\v\over m}\A\biggr)+
\biggl(g\pm {\pi\over mk}\biggr)\vp^2\biggr\}
\eqn\
$$
Let us compute the contribution of the second term:
$$
{1\over 2}\II \vec\nabla\wedge\biggl(\vec J+{\v\over m}\A\biggr)=
{1\over 2} r \int d\theta\, {\v\over m}{n\over r}={\pi\v\over m}n
\eqn\
$$
Therefore the total energy is:
$$
E=\pm {\pi\v\over m}n+\II\biggl\{{1\over 2m}|(D_x\pm iD_y)\phi|^2+
\Bigl(g\pm {\pi\over mk}\Bigr)\vp^2\biggr\}
\eqn\pietro
$$
{}From this expression we can learn several things. Firstly we note that the
energy $E$ is positive definite, see equation \ren. We will always study
the case where $g\!\geq\!{\pi\over mk}$.
\p
If one wants to relate \lagr\ to the Anyon mean field we
notice that the value of $g$ which reproduces the Anyon mean field energy
is $g\!=\!{\pi\over m}(1-{1\over k})$ (see Appendix A) and we see that
${\pi\over m}(1-{1\over k})\!\geq\!{\pi\over mk}$ with $k\!\geq\!2$.
\p
Therefore the integral on the r.h.s. of equation \pietro\ is positive or
zero. So we derive the inequality:
$$
E\geq {\pi\v\over m}|n|\geq {2\pi\over mk}\v={\cal E}
\eqn\bound
$$
So we have found that also the ``big'' deformations have energies over the
gap ${\cal E}$.
\p
Let us now study the special case $g\!=\!{\pi\over mk}$ and $n\!<\!0$,
where the bound \bound\ is saturated. In this case the equations simplify
because the energy becomes:
$$
E={\pi\v\over m}|n|+{1\over 2m}\II\Bigl\{|(D_x-iD_y)\phi|^2\Bigr\}
\eqn\
$$
which is minimal for:
$$
(D_x-iD_y)\phi=0.
\eqn\self
$$
Following Jackiw and Weinberg \jack\ we call equation \self\ self-dual
condition.
\p
Let us now solve the equation \self. In polar coordinates it is written:
$$
{\pa f\over \pa r}+{n\over r} f- \At f=0.
\eqn\polar
$$
If we introduce the auxiliary variable
$$
a=-n+r\At
\eqn\
$$
with the properties:
$$
\eqalign{
&a(0)=-n>0 \cr
&\lim_{r\to\infty}{a(r)}=0\cr}
\eqn\
$$
equation \polar\ is equivalent to the system:
$$
\left\{
\eqalign{&\pa_r f={1\over r} a f\cr
&\pa_r a=-{2\pi\over k}r\bigl(\v-f^2\bigr)\cr}\right.
\eqn\system
$$
This non-linear couple of differential equations has no analytic solution;
so we will study its asymptotic behaviours and then solve it numerically.
\p
As $r\!\rightarrow\!\infty$ the equations can be linearized defining
$F\!=\!(v\!-\!f)\!\rightarrow\! 0$ and we get an asymptotic solution in terms
of Bessel functions:
$$
\left\{
\eqalign{&F(r)=\alpha K_0\biggl(\sqrt{{4\pi\over k}}\, vr\biggr)\cr
&a(r)=\alpha \sqrt{{4\pi\over k}}\, r K_1\biggl(\sqrt{{4\pi\over k}}\,
vr\biggr)
\cr}\right.
\eqn\
$$
here
$$
\lim_{z\to\infty}{K_0(z)}=\sqrt{{\pi\over 2z}}\, e^{-z} \qquad\qquad
K_1(z)=-K'_0 (z)
\eqn\
$$
and $\alpha$ is some constant.
\p
For $r\!\rightarrow\!0$ we have:
$$
\left\{
\eqalign{&f(r)=A_n r^n+\cdots\cr
&a(r)=n-{\pi\over k}\v r^2+\cdots\cr} \right.
\eqn\
$$
$A_n$ being some constant.
\p
We solved the equation numerically (with DO2GAF-NAG Fortran Library Routine)
and verified the agreement of the solution with the previous discussion.
\p
In the case of $n\!>\!0$, the equations cannot be reduced to first order
and the bound cannot be saturated. We will study this case by a variational
method which, in agreement with \bound, confirms that the vortex energy for
$n\!=\!+|n|$ is in fact higher than that for $n\!=\!-|n|$.
\p
This fact should be expected since the opposite signs of $n$ are connected
by a parity transformation, but our Lagrangian density \lagr\ is not parity
invariant, so vortices that differ for the sign of $n$ cannot have the
same energy.

\parag{2.2.1.}{Currents}
\noindent
In this subsection we compute the electric current density and the total
current due to the vortex for $n\!<\!0$
\p
{}From \current\ using equation \polar\ we get:
$$
\left\{
\eqalign{&J_r =0\cr
&J_{\theta}=-{e\over m} f\pa_r f \cr}\right.
\eqn\
$$
The total current $I$ passing in the plane is equal to
$$
I=\int d\vec s\cdot\vec J=\int^{\infty}_0 dr\,J_{\theta}(r)=
-{e\over m}\int^{\infty}_0 dr\, f\pa_r f=
-{e\over m}\int^{\infty}_0 dr\, {d\over dr} f^2=
$$
$$
=-{e\over m}\Bigl[f^2(\infty)-f^2(0)\Bigr]=
-{e\over m}\v
\eqn\
$$
This current can be interpreted as the Hall current of a QHE!
\p
To see this fact let us consider the following Hamiltonian:
$$
H=|n|{\pi\v\over m}+{1\over 2m}\II\Bigl\{\dps +eB|\phi|^2\Bigl\}
\eqn\qhe
$$
where $B\!=\!-{2\pi\over ke}\vp$.
It is straightforward to prove that this Hamiltonian is equivalent to our
original Hamiltonian with $g\!=\!{\pi\over mk}$. In this form
we can interpret the second term in curly brackets in \qhe\ as an
interaction term between the electric potential $V\!=\!{1\over 2m} B$ and
the electric charge density $e|\phi|^2$. To such a potential will
correspond the electric field $\vec E\!=\!-\vec\nabla V$ which in polar
coordinates reads:
$$
\left\{
\eqalign{&E_r =-{\pa V\over\pa r}=
-{1\over 2m}{\pa B\over\pa r}=
-{2\pi\over mke}\,f\pa_r f=
-{2\pi\over ke^2}J_{\theta}\cr
&E_{\theta}=-{1\over r}{\pa V\over\pa\theta}=0\cr}\right.
\eqn\er
$$
{}From the first of the \er\ we read the value of the Hall conductivity
$\sigma\!=\!{J_\theta\over E_r}$:
$$
\sigma=k{e^2\over 2\pi}.
\eqn\
$$

\parag{2.2.2.}{Variational method}
\noindent
We have found exact vortex-like solutions obeying the self-dual condition
\self\ for $n\!<\!0$. In order to further investigate the general case we
try to find the vortices in a variational way. This can be done with
the following ansatz for $\AT$:
$$
r\AT_{\theta}(r)=n\Bigl[1-\Bigl(1+{p^2 r^2\over 2}\Bigr)\emr\Bigr]
\eqn\atansatz
$$
and minimizing the energy \toten\ with respect to the parameters $p$ and
$\omega$. Correspondently we get for $f(r)$, see equation \vortex, imposing
the condition $f(0)\!=\!0$ which yields the relation
$p^2\!=\!2\omega\!+\!{nk\v\over 2\pi}$ between the parameters,
$$
f^2(r)=\v-\Bigl(\v-{nk\over 2\pi}p^2\mr\Bigr)\emr
\eqn\
$$
In the minimization program we have taken
$|n|\!=\!1,\, g\!=\!{\pi\over m}\Bigl(1-{1\over k}\Bigr),\, k\!=\!2$ to compare
the result with the minimum find above, see equation \bound.
\p
For $n\!<\!0$ we have found $E_{min}\!=\!6.44\,{\v\over 2m}$ which is very
close ($6.44\!\simeq\!2\pi$) to the exact result, \ie the gap ${\cal E}$.
\p
For $n\!>\!0$ we have found $E_{min}\!=\!20.64\,{\v\over 2m}$ which is
quite above the gap.
\p
Notice that the variation in the number of the particles due to these
vortices is:
$$
\delta N=\II\delta\rho=\II\Bigl(f^2(r)-\v\Bigr)=
\II\Bigl[\Bigl({nk\over 2\pi}p^2\mr-\v\Bigr)\emr\Bigr]=nk
\eqn\
$$
So the total number of particles is preserved only if the total vorticity
is zero.
\goodbreak

\subchapter{2.3.}{Summary of the spectrum}
\noindent
We can conclude saying that all the excitations of the system lie over the
gap.
\p
Recalling equation \spec, with the correction coming from the electrostatic
interaction between excitations (see the comment after equation \spec), we see
that for small $p$ we have:
$$
E(p)\simeq{\cal E}+{e^2k\over 8\pi}|\vec p|
\eqn\
$$
For larger values of $p$ it is possible to have a roton excitation
\zhk\ \leezhang\ \leefisher\ \zhang\  which can be explained
as a Coulomb interaction between vortices.
\p
In fact, the Coulomb interaction gives an additional positive contribution
to the energy, namely the electrostatic energy due to the density
fluctuation of a charged fluid. This electrostatic energy is obviously
positive for any density fluctuation and in particular for the
vortex-antivortex configurations. Therefore, in conclusion, the roton part
of the spectrum will correspond to an energy higher than the lower bound of
a vortex-antivortex configuration, that is twice the gap ${\cal E}$. In
Fig. 1 we report the qualitative plot of the energy spectrum versus the
momentum. (In the case considered in reference \zhang\ the vortices are
assumed to appear in the lowest Landau level of a Hall system, whereas the
gap of the small deformations is due to excitations to the higher Landau
levels. Thus, in this case the vortices can have an energy less that the
gap. The case studied in reference \zhang\ would correspond in our
formalism to $g\!<\!{\pi\over mk}$. In our case instead we take
$g\!\geq\!{\pi\over mk}$, see the discussion of section 2.2).
\nobreak
\null
\nobreak
\vskip9truecm
\nobreak
\centerline{{\it Fig. 1}}
\vfill
\goodbreak

\newcount\equanumber	\equanumber=0
\def\chapterlabel{3}
\def\eqname#1{\relax\global\advance\equanumber by 1
  \xdef#1{{\rm(\chapterlabel.\number\equanumber)}}#1}

\def\eqn{\eqno\eqname}

\chapter{3.}{Chiral property}

In this section we study how the chirality properties of our superfluid
show up in the optical activity. Similar studies have appeared in the
literature for other kinds of $T$- and $P$-breaking theories, see in
particular ref. \ref\Wen. Let us consider an electromagnetic wave incoming
on a thin layer of superfluid, perpendicularly to the plane of the layer.
We begin the analysis studying how the electromagnetic potential couples
with our system on the plane. The small deformation Lagrangian, equation
\smdeflag, is modified adding the incoming electromagnetic potential $\A$:
$$
{\cal L}=2v\theta\vec\nabla\cdot\skew5\dot{\U}-
{1\over 2m}\v\biggl(\vec\nabla\theta-
{4\pi\v\over k}\skew5\tilde{\U}+e\A\biggr)^2-
{1\over 2m}(\triangle\U)^2-
4g\v(\vec\nabla\cdot\U)^2
\eqn\polag
$$
here $\skew5\tilde{\U}$ is the dual of $\U$ in the sense that
$\tilde{u}_i\!=\!\epsilon_{ij}u_j$. $\A$ is a two dimensional vector lying
in the plane. All the derivative operators act here on the two dimensional
coordinates on the plane.
\p
Performing the variation with respect to $\theta$ we find:
$$
{\delta{\cal L}\over\delta\theta}=
2v\vec\nabla\cdot\skew5\dot{\U}+
{\v\over m}\vec\nabla\cdot\biggl(\vec\nabla\theta-
{4\pi\v\over k}\skew5\tilde{\U}+e\A\biggr)=0
\eqn\
$$
{}From \irrot\ we have $\vec\nabla\cdot\skew5\tilde{\U}\!=\!0$.
Decomposing $\A\!=\!\A_L\!+\!\A_{\perp}$,
where by definition $\vec\nabla\cdot\A_{\perp}\!=\!0$, we have:
$$
\vec\nabla\theta=-{2m\over v}\skew5\dot{\U}-e\A_L
\eqn\
$$
Then the lagrangian \polag\ can be rewritten as:
$$
{\cal L}=-2v\skew5\dot{\U}\cdot\vec\nabla\theta-
{1\over 2m}\v\bigl(\vec\nabla\theta+e\A_L\bigr)^2-
{1\over 2m}\v\biggl(e\A_{\perp}-{4\pi\v\over k}\skew5\tilde{\U}\biggr)^2-
{1\over 2m}(\triangle\U)^2-
$$
$$
-4g\v(\vec\nabla\cdot\U)^2
\eqn\ppol
$$
Now we introduce the following parameterization
$\A\!=\!\vec\nabla\varphi\!+\!\skew5\tilde{\vec\nabla}\psi$, that is
$A_i\!=\!\pa_i\varphi\!+\!\epsilon_{ij}\pa_j\psi$. Thus equation \ppol\
becomes:
$$
{\cal L}=2m\skew5\dot{\U}^2+
2ve\skew5\dot{\U}\cdot\vec\nabla\varphi-
2m{\cal E}^2\U^2+
2v{\cal E}\U\cdot\vec\nabla\psi-
{e^2\v\over 2m}(\vec\nabla\psi)^2-
{1\over 2m}(\triangle\U)^2-
$$
$$
-4g\v(\vec\nabla\cdot\U)^2
\eqn\
$$
Taking $\U\!=\!\U_0e^{i(\omega t+\vec p\cdot\x)}$ we find:
$$
{\cal L}=2m\Bigl[\omega^2-{\cal E}^2-g(p)\Bigr]\U^2+
2v\U\cdot(i\omega\vec\nabla\varphi+{\cal E}\vec\nabla\psi)-
{e^2\v\over 2m}(\vec\nabla\psi)^2
\eqn\michela
$$
here $g(p)$ stands for the $p$-dependent terms in \spec.
\p
By performing the functional integration over
$\U$ and remembering that
$\vec\nabla\varphi={\vec p\cdot\A\over p^2}\,\vec p$ and\goodbreak\noindent
$\vec\nabla\psi={\epsilon_{ij}A_ip_j\over p^2}\,\vec p$,
we get a quadratic piece in $A$
$$
{\cal L}_{eff}=-{1\over 2}A^{\dag}\hat{\cal L}A
\eqn\
$$
where:
$$
\hat{\cal L}={e^2\v\over m}{1\over\omega^2-{\cal E}^2-g(p)}{1\over p^2}
\times (M_1+M_2)
\eqn\
$$
The matrices $M_1$ and $M_2$ are given by:
$$
M_1=\left(\matrix{
|ip_x\omega+p_y{\cal E}|^2&
(ip_x\omega+p_y{\cal E})^*(ip_y\omega-p_x{\cal E})\cr
(ip_y\omega-p_x{\cal E})^*(ip_x\omega+p_y{\cal E})&
|ip_y\omega-p_x{\cal E}|^2\cr}\right)
\eqn\
$$
$$
M_2={e^2\v\over mp^2}
\left(\matrix{
p_y^2&
-p_xp_y\cr
-p_xp_y&
p_x^2\cr}\right)
\eqn\
$$
Here we have introduced the following notation:
$$
A=\left(\matrix{A_x\cr
A_y\cr}\right)
\eqn\
$$
${\cal L}_{eff}$ is the contribution to the total lagrangian coming from
the interaction between the incoming wave and the two dimensional chiral
fluid. Now we suppose $\A$ propagating in the $z$ direction and the plane
situated at $z\!=\!0$; if we write $A\!=\!A(z)e^{i(\omega t+\vec p\cdot\x)}$,
with $A(z)$ discussed below, the total lagrangian is:
$$
{\cal L}(A)={1\over 2}\omega^2 a^2-
{1\over 2}\bigl[p^2A^2+(\pa_zA)^2\bigr]-
{1\over 2}A^{\dag}\,\hat{\cal L}\,A\,\delta(z)
\eqn\llint
$$
from equation \llint\ we extract the equation of motion for $A$:
$$
\pa_z^2 A+(\omega^2-p^2)A-\hat{\cal L}\,A\,\delta(z)=0
\eqn\
$$
For $A(z)$ we have:
$$
A(z)=\cases{
\alpha_{-}e^{ikz}+\beta_{-}e^{-ikz}&if $z<0$\cr
\alpha_{+}e^{ikz}&if $z>0$\cr}
\eqn\
$$
Imposing continuity at $z\!=\!0$ for $A(z)$ and its derivative, we get:
$$
\alpha_{+}={2ik\over 2ik-\hat{\cal L}}\alpha_{-},
\beta_{-}=\alpha_{+}-\alpha_{-}={\hat{\cal L}\over 2ik-\hat{\cal L}}\alpha_{-}
\eqn\retrans
$$
{}From equation \retrans\ we can read the matrices of reflection and
transmission:
$$
R={{\hat{\cal L}\over 2ik}\over 1-{\hat{\cal L}\over 2ik}} \qquad
T={1\over 1-{\hat{\cal L}\over 2ik}}
\eqn\
$$
Notice that $R\!=\!T\!-\!1$ and $|T|^2+|R|^2\!=\!1$ as it should.
\p
In particular we are interested in the coefficient of transmission in the
limit when $\omega\!\to\!{\cal E}$. In this case only $M_1$ in equation
\llint\ is important. If we compute the inverse of the matrix
$1-{1\over 2ik}\hat{\cal L}$ we get, taking for simplicity $\vec p=(0,p)$:
$$
T={1\over 1-2i(q+s)-2qs}
\left(\matrix
{1-iq&-q\cr
q&1-i(q+2s)\cr}
\right)+O(1/q)
\eqn\
$$
here we have set
$q=-{1\over 2k}{e^2\v\over m}{{\cal E}^2\over\omega^2-{\cal E}^2}$ and
$s=-{1\over 2}{1\over 2k}{e^2\v\over m}$.
\p
The limit $\omega\!\to\!{\cal E}$ corresponds to $q\!\to\!\infty$ that is:
$$
T={1\over 2}{i\over i+s}
\left(\matrix
{1&-i\cr
i&1\cr}
\right)
\eqn\
$$
Now let us consider the following parameterization:
$$
\eqalign{
A_x&=Re\bigl(\cos\theta e^{-i\omega t}\bigr)=\cos\theta\cos\omega t \cr
A_y&=Re\bigl(\sin\theta e^{i\varphi}e^{-i\omega t}\bigr)=
\sin\theta\cos(\varphi-\omega t) \cr}
\eqn\
$$
we can see that the circular polarization is for $\theta\!=
\!{\pi\over 4}$ and $\varphi\!=\!{\pi\over 2}$, so that:
$$
A_x={\sqrt{2}\over 2}\cos\omega t \qquad A_y={\sqrt{2}\over 2}\sin\omega t
\eqn\
$$
in our complex formalism  this corresponds to:
$$
A=\left(\matrix{\cos\theta\cr
\sin\theta e^{i\varphi}\cr}
\right)e^{-i\omega t}=
{\sqrt{2}\over 2}\left(\matrix{1\cr
i\cr}\right)e^{-i\omega t}
\eqn\
$$
Thus we see that $T$ projects a state in a circularly polarized one,
that is we can write:
$$
T={1\over 2}\,{i\over i+s}\,|c\rangle\langle c|
\eqn\
$$
where
$$
|c\rangle=\left(\matrix{1\cr
i\cr}\right)
\eqn\
$$
That is, at the resonance $\omega\!\to\!{\cal E}$ our chiral planar system
behaves like a perfect polarizer.
\p
Finally note that if $\omega\!\to\!\infty$ we get:
$$
\hat{\cal L}\to {e^2\v\over m}
\left(\matrix{
1&0\cr
0&1\cr}
\right)
\eqn\
$$
so we have lost any polarizing effect.
\p
If conversely we take $\omega\!\to\!0$ and $p\!\to\!0$ we get
$\hat{\cal L}=0$, so there is no more coupling between the electromagnetic
wave and the planar system.

\bigskip
\bigskip
\bigskip

\newcount\equanumber	\equanumber=0
\def\chapterlabel{4}
\def\eqname#1{\relax\global\advance\equanumber by 1
  \xdef#1{{\rm(\chapterlabel.\number\equanumber)}}#1}

\def\eqn{\eqno\eqname}

\chapter{4.}{Coupling with an external magnetic field}

\subchapter{4.1.}{Magnetic field orthogonal to the layers}

\parag{4.1.1.}{Meissner effect}
\noindent
In this section we test our chiral charged superfluid with an external
magnetic field, and study its Meissner effect.
\p
Up to now we have been considering a two dimensional system; now we are
going to study an effect which is essentially three-dimensional. Therefore
we suppose to have a multilayered bulk of many two-dimensional thin films
separated by a spacing $d$ (for a review on the properties of the
layered superconductors see, for example \ref\revBula). We further suppose
that at the edge of the bulk there is a uniform, constant magnetic field
orthogonal to the layers' plane as in Fig. 2a.
\null
\vskip8truecm
\centerline{\it Fig. 2}
\null
\medskip
\noindent
In this three dimensional system we must take into account also an
electrostatic contribution, (see the comment after equation \spec) and
which will play an essential r\^ole here. This contribution is essentially
due to an electric field $\E$ which comes from the fluctuations of the
charged matter and obeys the Maxwell equation:
$$
\vec\nabla\cdot\E=e\,\delta\rho^{(3)}
\eqn\
$$
here $\delta\rho^{(3)}\!=\!{\delta\rho\over d}$, is the three dimensional
density.
\p
In other words if there is a fluctuation $\delta\rho$ of matter the system
will not be in electrostatic equilibrium anymore, for there will be some zones
where there is lack of charged matter and others where there is abundance:
$\E$ is the electric field resulting from this non-equilibrium situation.
\p
So let us consider the following three-dimensional Hamiltonian density (in
this section we have redefined $\skew6\vec{\AT}\!\to\!\ACS$ to distinguish
from $\Aem$):
$$
{\cal H}={1\over 2md}\dps+
{1\over 2}(\B^2+\E^2)+
{g\over d}(\rho-\v)^2
\eqn\
$$
here
$$
\B=\vec\nabla\wedge\Aem \qquad
\E=-\vec\nabla A_0^{em} \qquad
\D=\vec\nabla-i\A^{CS}-ie\Aem
\eqn\
$$
Notice that the following equations hold:
$$
\cases{
\vec\nabla\cdot\E={e\over d}\delta\rho\cr
\vec\nabla\wedge\E=0\cr}
\qquad\qquad
\cases{
\vec\nabla\cdot\ACS=0\cr
\vec\nabla\wedge\ACS={2\pi\over k}\delta\rho\cr}
\eqn\dual
$$
here $\E$ is a three-dimensional vector whereas $\ACS$ is two-dimensional.
If we suppose that the matter distribution is constant in the $z$
direction, which means that we have exactly the same matter distribution in
every layer, then we have no electric field orthogonal to the $x$-$y$
layers' plane. With this assumption equations \dual\ tell that $\E$
and $\ACS$ are dual two-dimensional vectors, \ie:
$$
E_i={ke\over 2\pi d}\epsilon_{ij}A_j^{CS}
\eqn\
$$
So we can write the hamiltonian as follows:
$$
{H\over L_z}=\int dxdy\,\biggl\{{1\over 2md}\dps+
{1\over 2}\biggl[\bigl(\vec\nabla\wedge\Aem\bigr)^2+
{k^2 e^2\over 4\pi^2 d^2}(\ACS)^2\biggr]+
{g\over d}(\rho-\v)^2\biggr\}
\eqn\hsulz
$$
here we have performed the integration in $z$.
\p
At this point we make one further assumption: we suppose that the external
magnetic field is constant in $y$, which is the direction parallel to the
edge of the bulk, so we can take $\Aem(x)$ in the $y$ direction:
$$
\Aem=(0,\aem(x),0)
\eqn\
$$
furthermore we still have the freedom to take also $\ACS(x)$ in the $y$
direction:
$$
\ACS=(0,\aCS(x),0)
\eqn\
$$
In the gauge:
$$
\vec\nabla\!\cdot\!\Aem\!\!=\!\!\vec\nabla\!\cdot\!\ACS\!\!=\!\!0
\eqn\
$$
$\aem$ and $\aCS$ depend only on $x$. With these assumptions all quantities
in \hsulz\ depend only on $x$, so we can perform the integration in $y$ and
get:
$$
{H\over L_y L_z}=\int_0^{\infty}dx\,\biggl\{
{1\over 2md}\Bigl[|\pa_x\phi|^2+|e\aem+\aCS|^2\rho\Bigr]+
{1\over 2}(\pa_x \aem)^2+
$$
$$
+{k^2 e^2\over 8\pi^2 d^2}(\aCS)^2+
{g\over d}\bigl(\rho-\v\bigr)^2\biggr\}
\eqn\hsulylz
$$
$L_y$ being the length of the edge of the bulk.
\medskip
Now, before going into computational details, we want to spend some time
analyzing qualitatively the reason why we expect a Meissner effect. To this
end we will make some simplificatory assumptions. The first is to suppose
that some external device is keeping constant the magnetic flux
$\Phi_0=L_yl_x B$, where $L_y$ is the length of the edge of the sample and
$l_x$ is the penetration length of the magnetic field, so that
$$
Bl_x={\Phi_0\over L_y}=\varphi_0
\eqn\givenflux
$$
is given to the system from outside.
\p
The second assumption we make is to suppose $B$ constant for
$0\!<\!x\!<\!l_x$ and zero for $x\!>\!l_x$ whereas actually it is
exponentially decreasing.
\p
The third assumption we make is to take a constant value for the density
$\rho\!=\!\v$.
\p
So we have for $0\!<\!x\!<\!l_x$:
$$
B={\varphi_0\over l_x} \qquad \aem(x)={x\over l_x}\varphi_0
\eqn\marta
$$
For $x\!>\!l_x$, $\aem$ is constant and it is possible to cancel it in the
covariant $y$-derivative with the phase of $\phi$. Thus, the integration
runs from $0$ to $l_x$.
\p
To test the meaningfulness of what we are doing let us see what happens for
the well known case of the standard superconductor, \ie with $\ACS\!=\!0$;
the hamiltonian \hsulylz\ becomes:
$$
{H\over L_y L_z}=\int_0^{l_x}dx\,\biggl\{
{1\over 2}\B^2+{1\over 2md}e^2(\aem)^2 \v\biggr\}=
{1\over 2}{\varphi_0^2\over l_x}+{1\over 6md}e^2\v \varphi_0^2 l_x
\eqn\
$$
Minimizing $H$ with respect to $l_x$ we find $l_x=\sqrt{{3md\over e^2\v}}$
which is indeed of the same order as the standard value of the penetration
length for type II superconductors:
$$
\lambda=\sqrt{{m\over e^2\rho^{(3)}}}
\eqn\lx
$$
Notice that in what we have done a fundamental r\^ole is played by the
term quadratic in $\aem$, the ``mass term'' for the electromagnetic field.
\p
In our case, with $\ACS\!\ne\!0$, this effect could be ruined by the
possible cancellation $\ACS\!=\!-e\Aem$ but, notice, we have also
the electrostatic interaction term quadratic in $\ACS$ which now plays the
dominant r\^ole. We have:
$$
{H\over L_y L_z}=\int_0^{l_x}dx\,\biggl\{
{1\over 2}\B^2+{1\over 2}\E^2+{1\over 2md}(e\aem+\aCS)^2 \v\biggr\}=
{1\over 2}{\varphi_0^2\over l_x}+{k^2 e^4\over 24\pi^2 d^2}\varphi_0^2 l_x
\eqn\sofri
$$
Minimizing with respect to $l_x$ we get $l_x={2\pi d \sqrt{3}\over ke^2}$
we see that we get a {\it finite} penetration length, so we expect to have
Meissner effect.
\medskip
Now we turn back to hamiltonian \hsulylz\ and analyze it more
quantitatively in the framework of the small deformation approach
introduced in the previous chapter.
\p
So we have, recalling the basic definitions, see equations
\param, \fluc\ and \afluc:
$$
\phi=f \qquad \pa_x\phi(x)=\pa_x^2 u(x)
\eqn\
$$
$$
\delta\rho=2v\pa_x u(x)
\eqn\
$$
$$
\aCS={4\pi v\over k}u \qquad g={\pi\over m}\biggl(1-{1\over k}\biggr)
\eqn\
$$
Notice that in this one-dimensional case $\phi$ is real, so there is no
phase.
\p
The hamiltonian becomes:
$$
{H\over L_y L_z}={1\over d}\int_0^{\infty}dx\,\biggl\{
{1\over 2m}\biggl[(\pa_x^2 u)^2+\biggl(e\aem+{4\pi v\over
k}u\biggr)^2\v\biggr]+
$$
$$
+{d\over 2}(\pa_x\aem)^2+{2\v e^2\over d}u^2+
{4\pi\v\over m}\biggl(1-{1\over k}\biggr)(\pa_x u)^2\biggr\}
\eqn\
$$
We now make the following ansatz:
$$
u(x)=u_0e^{-\lambda x} \qquad \aem(x)=A_0e^{-\lambda x}
\eqn\ansatz
$$
here $u_0$ and $A_0$ are the values at the edge of the bulk.
The resulting equations for $u(x)$ and $\aem(x)$ can be written in the form:
$$
\left(\matrix{
{\lambda^4\over 2m}-{4\pi\v\over m}(1-{1\over k})\lambda^2+
{8\pi^2 v^4\over mk^2}+{2e^2\v\over d}&
{2\pi v^3 e\over mk}\cr
{2\pi v^3 e\over mk}&
{e^2\v\over 2m}-{d\over 2}\lambda^2\cr}\right)
\left(\matrix{u\cr\aem\cr}\right)=0
\eqn\
$$
Putting to zero the determinant of this matrix we get:
$$
{d\over 4m}\lambda^6-
\biggl[{2\pi\v d\over m}\biggl(1-{1\over k}\biggr)+
{e^2\v\over 4m^2}\biggr]\lambda^4+
\biggl[{8\pi v^4 d\over 2mk^2}+e^2\v+
{2\pi e^2 v^4\over m^2}\biggl(1-{1\over k}\biggr)\biggr]\lambda^2-
$$
$$
-{e^4v^4\over md}=0
\eqn\terzogrado
$$
\goodbreak
If we take the following typical values for the parameters (as an order of
magnitude we take $m$ to be the electron mass):
$$
d=1{\rm\AA} \qquad e^2={4\pi\over 137} \qquad \v=4\cdot 10^{-3}{\rm\AA^{-2}}
\qquad m=250{\rm\AA^{-1}} \qquad k=2
\eqn\parameters
$$
we get the solutions:
$$
\lambda_1=1.21\cdot 10^{-3} \qquad
\lambda_{2,3}=0.56\pm i0.54\equiv\alpha\pm i\beta
\eqn\pian
$$
We can approximately compute analytically the smallest eigenvalue
$\lambda_1$ by rewriting equation \terzogrado\ in an approximate form
taking only the leading terms:
$$
\lambda^2\biggl\{{d\over 4m}\lambda^4-
\biggl[{2\pi\v d\over m}\biggl(1-{1\over k}\biggr)\biggr]\lambda^2+
e^2\v\biggr\}={e^4v^4\over md}
\eqn\
$$
for small $\lambda^2$ we get the solution:
$$
\lambda^2={e^4v^4\over md}{1\over e^2\v}={e^2\v\over md}
\eqn\penlength
$$
which, compared with equation \lx, is exactly the expression for the
inverse of the square of the standard penetration length, and, corresponds
numerically to $\lambda_1$ in equation \pian.
\p
Let us note that by solving \terzogrado\ in the limit of very small $e^2$ we
would find:
$$
\lambda={ke^2\over\sqrt{4\pi}d}\bigl(1+O(e^2)\bigr)
\eqn\
$$
which, apart for inessential numerical factors, is equal to the inverse of
the $l_x$ found minimizing \sofri. Indeed the qualitative analysis leading to
\sofri\ was meant in the limit of a very small electrostatic interaction.
Actually this is not the case. For instance, with the values given in
\parameters\ we get ${8\pi v^4d\over 2mk^2}\!\ll\!e^2\v$, and
\penlength\ follows.
\p
The general solution for $u(x)$ is the linear combination:
$$
u(x)=u_1e^{-\lambda_1 x}+u_2e^{-\lambda_2 x}+u_3e^{-\lambda_3 x}
\eqn\
$$
Imposing the reality condition $u^*(x)\!=\!u(x)$ we can write:
$$
u(x)=u_1e^{-\lambda_1 x}-e^{-\alpha x}(u_1\cos\beta x-w\sin\beta x)
\eqn\
$$
giving for $\delta\rho$:
$$
\delta\rho=2v\pa u= 2v\Bigl[-\lambda_1u_1e^{-\lambda_1 x}+
e^{-\alpha x}\bigl((\alpha u_1+\beta w)\cos\beta x+
(\beta u_1-\alpha w)\sin\beta x\bigr)\Bigr]
\eqn\
$$
Requiring $\rho$ to vanish at the edge, \ie $\delta\rho(0)\!=\!-\v$ we can
determine the value of $w$:
$$
w=-{v\over 2\beta}-{u_1\over\beta}(\alpha-\lambda_1)
\eqn\
$$
The following relation holds between $\aem(x)$ and $u(x)$:
$$
\aem(x)={4\pi v^3e\over mkd}{1\over \lambda^2-{e^2\v\over md}}u(x)
\eqn\
$$
We then compute $B\!=\!\pa_x\aem$.
\p
Imposing $B(0)\!=\!B_0$ we get a value for $u_1$, the only parameter still
undetermined. If we substitute the values of the parameters we see that the
exponential behaviour of $B$ is controlled by the first eigenvalue, \ie the
penetration length is ${1\over\lambda_1}=826.45 {\rm\AA}$. We can compare
this value to the numerical value of the standard penetration length
\lx\ which is $825.57 {\rm\AA}$.
\p
It is seen numerically that the behaviour of $\delta\rho$ is controlled
by the other eigenvalues, \ie the coherence length is
${1\over\alpha}=1.78{\rm\AA}$.
\p
Notice that the penetration length is about 400 times the coherence length,
so our system behaves like a type II superconductor.

\parag{4.1.2.}{Vortices}
\noindent
In this section we study, with a variational method, the vortices in
pre\-sen\-ce of an external electromagnetic field orthogonal to the layers'
plane.
\p
These vortices have an origin different from the vortices studied in
section 1.2. Those were originated by fluctuations of the CS magnetic
field, \ie of the matter density from the mean value. The present ones are
instead the standard well known vortex configurations of superconductors of
type II between the two critical temperatures. They consist of small
regions of the specimen of normal behaviour, all surrounded by a
superconducting region, where the external magnetic field penetrates
completely and uniformly. There are superconducting currents flowing around
the vortices. Furthermore there is a penetration of the magnetic field from
the vortex region to the surrounding superconducting region (Meissner
effect).
\p
As we saw in the previous section the hamiltonian is:
$$
{H\over L_z}=\II\biggl\{{1\over 2md}\dps+
{1\over 2}\B^2+{k^2e^2\over 8\pi^2 d^2}(\aCS)^2+{g\over d}(\rho-\v)^2\biggr\}
\eqn\hami
$$
where we have included the electrostatic effect, that is the third term on
the r.h.s., as in equation \hsulz.
\p
We make the following ansatz:
$$
\eqalign{
&\phi=f(r)e^{in\theta}\cr
&er\aem_{\theta}=n\Bigl(1-e^{-\mu r^2}\Bigr) \qquad \aem_{r}=0\cr
&r\aCS_{\theta}={2\pi\over k}\int_0^r dr'r'\bigl[f^2(r')-\v\bigr]=
{\pi\over k}\v r^2 \emr \qquad \aCS_{r}=0\cr}
\eqn\voransatz
$$
from the last one we easily get:
$$
f^2(r)=\v-\v(1-\mr)\emr
\eqn\
$$
Notice that our ansatz is such that we have quantization of the flux of the
external magnetic field, differently from the vortices studied previously
when it was the flux of the CS magnetic field to be quantized.
Notice also that since the vorticity is no longer connected with the CS
field now, differently from the vortices studied before, it is possible
to have an isolated vortex or antivortex since
$\int_0^{\infty}drr\bigl[\v-f^2(r)\bigr]=0$
as can be checked from the last of equations \voransatz.
\p
If one substitutes in \hami\ and minimizes, numerically,  with respect to
$\lambda$ and $\omega$, with the usual values of the parameters
\parameters\ one gets:
$$
\omega=0.101{\rm\AA^{-2}} \qquad\qquad \mu=7.336\cdot 10^{-7}{\rm\AA^{-2}}
\eqn\
$$
for the case of the vortex ($n\!=\!1$)
and
$$
\omega=0.117{\rm\AA^{-2}} \qquad\qquad \mu=7.337\cdot 10^{-7}{\rm\AA^{-2}}
\eqn\
$$
for the case of the antivortex ($n\!=\!-1$).
\p
{}From these results we get the values of the dimensions of the vortex and of
the fluctuations of the magnetic field:
$$
{1\over\sqrt{\omega}}=3.147{\rm\AA} \qquad\qquad
{1\over\sqrt{\mu}}=1167.54{\rm\AA}
\eqn\
$$
for the vortex, and
$$
{1\over\sqrt{\omega}}=2.924{\rm\AA} \qquad\qquad
{1\over\sqrt{\mu}}=1167.46{\rm\AA}
\eqn\
$$
for the antivortex.
Notice that for the vortex and for the antivortex we have different
results, although very close to each other, as it should have been expected
since our system is chiral.
\p
We see that again the behaviour is that typical for a type II superconductor.

\parag{4.1.3.}{Layered structure of the Vortices}
\noindent
We can also study the properties of these vortices due to the fact that the
superconducting material is composed of a stack of many not-strongly-coupled
layers. We follow reference \ref\Clem\ where a three dimensional vortex is
built up superposing a stack of two dimensional vortices in the case of a
standard high $T_c$ superconductor. In that paper it is explicitly
computed, as a first step, the magnetic field, $\vec b$,  produced by a
single layer and then, using this result, the whole stack contribution is
computed. Here we will only show that in our case we can reach, under
reasonable assumptions, the same first step, and then just state the final
results.
\p
The problem has a cylindrical symmetry therefore we will use cylindrical
polar coordinates $(r, \theta, z)$.
The supercurrent flowing in our single layer is:
$$
\vec J=(0,J_{\theta},0) \qquad\qquad
J_{\theta}=J_{\theta}(r,z)=K_{\theta}(r)\delta(z)
\eqn\
$$
{}From Amp\`ere's law we get:
$$
\vec\nabla\wedge\vec b=\vec J \Rightarrow
\cases{
\pa_zb_r-\pa_rb_z=K_{\theta}(r)\delta(z)\cr
\pa_zb_{\theta}=0\cr
{1\over r}\pa_r(rb_{\theta})=0\cr}
\eqn\ampere
$$
from the last two we get $b_{\theta}\!=\!0$.
\p
For $z\!\neq\!0$ equation $\vec\nabla\wedge\vec b\!=\!0$ is solved by:
$$
A_{\theta}(r,z)=\int^{\infty}_0\, dqA_0(q)J_1(rq)e^{-q|z|} \qquad\qquad
A_r=A_z=0
\eqn\azeta
$$
notice that $\vec\nabla\cdot\vec A\!=\!0$. $A_0(q)$ is to be determined.
\p
Now from the first of \ampere, solved for $z\!=\!0$, and \azeta\ we can get:
$$
K_{\theta}(r)=b_r(r,0^{+})-b_r(r,0^{-})=2\int^{\infty}_0\, dqqA_0(q)J_1(rq)
\eqn\
$$
For our vortex we know:
$$
K_{\theta}=
{e\over m}\Bigl({n\over r}-e\aem_{\theta}-\aCS_{\theta}\Bigr)f^2(r)
\eqn\
$$
therefore, using \azeta:
$$
\int^{\infty}_0\, dqA_0(q)J_1(rq)\biggl[2q+{e^2\over m}f^2(r)\biggr]=
{e\over m}f^2(r)\Bigl({n\over r}-\aCS_{\theta}\Bigr)
\eqn\elena
$$
Now if the superconductor is of type II \ie if the dimensions of the vortex
are much smaller than the penetration length of $\vec b$, then it is
sensible to take $f^2(r)$ equal to its mean value $\v$ and
$\aCS_{\theta}\!=\!0$. Within this approximation \elena\ becomes:
$$
\int^{\infty}_0\, dqA_0(q)J_1(rq)\biggl(2q+{e^2\v\over m}\biggr)=
{e\v\over m}\Bigl({n\over r}\Bigr)
\eqn\
$$
Using the orthogonality property of the Bessel functions
$\int_0^{\infty}\, drrJ_1(rq)J_1(rq')={1\over q}\delta(q-q')$ we get:
$$
A_0(q)={en\v\over m}\biggl(2q+{e^2\v\over m}\biggr)^{-1}=
{n/e\over 1+\Lambda q} \qquad\qquad \Lambda={2m\over e^2\v}
\eqn\
$$
$\lambda$ is related to the nominal penetration length
$\lambda^2={m\over e^2\rho^{(3)}}$ by the relation:
$$
\Lambda={2\lambda^2\over d}
\eqn\
$$
and can be considered as the two dimensional penetration length.
\p
One can then compute \azeta\ and verify that the magnetic field decays with
the penetration length $\Lambda$.
\p
These results are exactly the same obtained in \Clem. We will not reproduce
here all the computations that can be found in \Clem, but just state the
main results.
\p
Having studied what happens with a single layer we possess the building
block to all the multilayered system superposing the entire stack of two
dimensional vortices.
\p
Then one can study the binding energy between vortices in different
layers and find that thermal excitation breaks up the stack above a
transition temperature corresponding to the Kosterlitz-Thouless temperature
for a bidimensional system, see \ref\Thouless\ \ref\Beasley.
\goodbreak
\subchapter{4.2.}{Magnetic field parallel to the layers}

\parag{4.2.1.}{Meissner effect}
\noindent
In this section we consider a different physical problem with $\B$ in the
plane of the layers. This problem, for the case of standard high $T_c$
superconductors, has been studied in many paper by J.R. Clem and
collaborators \Coffey\ \ref\Hao\ \ref\Bulaevskii\ \ref\revClem.
In our analysis we choose $\B$ pointing in the $x$-direction (see Fig. 2b),
and study the penetration length in $z$, $l_z$, supposing uniformity along
$y$. Let us start from the following Hamiltonian:
$$
{H\over L_y}=\int dxdz\biggl\{{1\over 2md}\Bigl|\Bigl(\vec\nabla-ie\Aem-
i\ACS\Bigr)\phi\Bigr|^2+{1\over 2}\B^2+g(\rho-\v)^2\biggr\}-
$$
$$
-{e^2\over 2}\int dxdx'dzdz'\biggl\{{\delta\rho(x',z')\over d}
\log\biggl[{(x-x')^2+(z-z')^2\over\Lambda^2}\biggr]{\delta\rho(x,z)\over d}
\biggr\}
\eqn\
$$
here we have introduced the electrostatic interaction between the
fluctuations of the charged matter which, due to uniformity in the
$y$-direction, is in fact the electrostatic interaction of a
two-dimensional distribution of charge, which is known to be logarithmic;
$\Lambda$ is a convenient dimensional constant which, since
$\int\delta\rho\!=\!0$, can take an arbitrary value. In the gauge where
$\Aem$ and $\ACS$ are both in the $y$-direction
$\Aem\!=\!(0,\aem(x,z),0),\,\ACS\!=\!(0,\aCS(x,z),0)$ we can rewrite the
Hamiltonian as:
$$
{H\over L_y}=\int dxdz\biggl\{{1\over 2md}\Bigl(|\pa_x\phi|^2+
{md^2j_0\over e}|\pa_z\phi|^2\Bigr)+{1\over 2md}\Bigl|e\aem+\aCS\Bigr|^2\rho+
{1\over 2}(\pa_z\aem_y)^2+
$$
$$
+g(\rho-\v)^2\biggr\}-{e^2\over 2d^2}\int dxdx'dzdz'\biggl\{\delta\rho(x',z')
\log\biggl[{(x-x')^2+(z-z')^2\over\Lambda^2}\biggr]\delta\rho(x,z)
\biggr\}
\eqn\paola
$$
here $j_0$ is a constant depending on the intrinsic features of the
material and measures the Josephson coupling between neighbouring layers.
\p
Similarly to what we have done for the field orthogonal to the layers we
consider the possible cancellation
$\aCS(x,z)\!=\!-\aem(z)\bigl[\theta(x)\theta(L_x-x)\bigr]$, $L_x$ being the
length of the $x$-edge of our sample, and $\theta$ is the step function.
Thus:
$$
\delta\rho(x,z)={k\over 2\pi}\pa_x\aCS(x,z)=
-{ke\over 2\pi}\aem(z)\bigl[\delta(x)-\delta(L_x-x)\bigr]
\eqn\
$$
The first and the last term in the first integral of \paola\ are edge effects
giving a contribution which is very small compared with the rest since, going
through the computation, one can see that they are suppressed respectively by
factors $(L_x)^{-1}$ and $(L_x)^{-1/3}$ So in the following we will skip them.
Thus:
$$
{H\over L_y}={L_x\over 2}\int dz(\pa_z\aem_y)^2
+{k^2e^4\over 4\pi^2d^2}\int dzdz'\biggl\{\aem(z)
\log\biggl[1+{L_x^2\over (z-z')^2}\biggr]\aem(z')\biggr\}
\eqn\
$$
Like we have done in the previous section (compare with equation \sofri), we
fix the total flux $\Phi_0\!=\!BL_yl_z$ and define
$\varphi_0={\Phi_0\over L_y}$ such that for $0\!<\!z\!<\!l_z$:
$$
B={\varphi_0\over l_z} \qquad\qquad \aem=-{\varphi_0\over l_z}z
\eqn\uani
$$
and the integration is from $0$ to $l_z$.
\p
Therefore we get:
$$
{H\over\varphi_0^2 L_y}={L_x\over 2l_z}
+{k^2e^4\over 4\pi^2d^2}{1\over l_z^2}\int_0^{l_z} dzdz'
\biggl\{zz'\log\biggl[1+{L_x^2\over (z-z')^2}\biggr]\biggr\}=
$$
$$
={L_x\over 2l_z}+{k^2e^4\over 16\pi^2}{l_z^2\over d^2}
\biggl[\log{L_x^2\over l_z^2}+\cdots\biggr]
\eqn\
$$
We assumed $L_x/l_z$ large to expand the logarithm.
\p
Minimizing $H$ with respect to $l_z$ we get:
$$
l_z\simeq \biggl({3\pi^2\over k^2e^4}\biggr)^{1/3}\cdot d\cdot
\biggl({L_x\over d}\biggr)^{1/3}\biggl[\log\biggl({L_x\over d}\biggr)+\cdots
\biggr]^{-1/3}
\eqn\iengo
$$
Therefore we have a ``quasi Meissner effect'' in the sense that even if
$l_z\!\to\!\infty$ for $L_x\!\to\!\infty$ we have ${l_z\over L_x}\!\to\!0$.
The result of equation \iengo\ appears to be quite peculiar of the chiral
superconductor studied in this paper.
\p
In fact, the non-chiral superconductors would show ordinary Meissner effect
for the geometry considered in this section (at least for not too small
Josephson coupling between neighbouring layers): that is, magnetic fields
parallel to the layers would penetrate in the direction orthogonal to the
layers (say the $z$-direction) for a finite distance $l_z$, of the order of
the standard penetration length.
\p
The chiral superconductor studied in this paper would instead penetrate
much more, for a distance $l_z$ as given in equation \iengo. This result
holds for arbitrary Josephson coupling between neighbouring layers.
\p
(Let us also mention that, in the case of very small Josephson coupling it
is possible to imagine another configuration possibly giving a large
penetration in $z$. This configuration is not peculiar of the chiral
superconductor studied here. We discuss this configuration in Appendix B).
\p
Finally we review a general argument for multilayer superconductors
indicating that the magnetic field parallel to the layers can penetrate
easily in the interlayer spacing, therefore in the direction parallel to
the layers (say in the $y$-direction, for a field along $x$), even in the
case in which $l_z$ is finite. How much the corresponding $l_y$ is large,
depends on how small is the interlayer Josephson coupling. This effect
would be basically the same for chiral and non-chiral superconductors. In
order to study this penetration in the direction parallel to the layers we
follow the study of ref. \Coffey, considering a possible vortex
configuration inside the material. From this configuration we will infer
the penetration properties. A crucial fact for this study is taking into
account the Josephson coupling between neighbouring layers
\Coffey\ \ref\Doniach. This study has already been worked out in
\Coffey\ for the case of a standard high $T_c$ superconductor, here we
merely rephrase that paper for our case. The main fact is that one has to
add to the supercurrent density:
$$
\vec J={e\v\over m}\bigl(\vec\nabla\gamma-e\A\bigr)
\eqn\
$$
(here $\gamma$ is the phase of $\phi$, equal to $in\theta$ for the vortex
configuration), the Josephson current flowing between to neighbouring layers,
say the $n$-th and the $(n\!+\!1)$-th, proportional to the sine of the gauge
invariant difference of phase of $\phi$ between the layers:
$$
j_z=j_0\sin(\Delta\gamma_n) \qquad\qquad
\Delta\gamma_n=\gamma_{n+1}-\gamma_n+e\int_{n}^{n+1}d\vec l\cdot\A
\eqn\
$$
$j_0$ is a constant depending on the material.
\p
Solving the Amp\`ere equation one finds \Coffey:
$$
b(\tilde r)={1\over el_z\lambda_J}K_0(\tilde r) \qquad \qquad
\tilde r=\sqrt{{y^2\over \lambda_J^2}+{z^2\over l_z^2}}
\eqn\
$$
with $\lambda_J^2\!=\!{1\over edj_0}$;
here $d$ is the stack periodicity of the layers.
\p
This result tells essentially that $\vec B$ penetrates differently along
$y$, \ie parallel to the layers, and along $z$, \ie orthogonally, in other
words the vortex has an ``elliptic symmetry''. Notice that in reference
\Coffey\ the decay length along the $y$-axis is called $\lambda_z$ instead
of $\lambda_J$; the reason is that this decay length is essentially due to the
screening currents which are pointing in the orthogonal direction \ie the
$z$-direction. Similar argument goes for the decay length along the
$z$-axis which in \Coffey\ is called $\lambda_y$.
\p
In conclusion, in both cases of ordinary and chiral superconductors the
magnetic field would easily penetrate along the interlayer spacing.
Instead, the penetration across the layers of a magnetic field parallel to
them will be finite for an ordinary superconductor (for not too small
Josephson coupling), whereas it will grow with a fractional power of the
sample's dimension for the chiral superconductor.

\vfill
\eject

\newcount\equanumber	\equanumber=0
\def\chapterlabel{A}
\def\eqname#1{\relax\global\advance\equanumber by 1
  \xdef#1{{\rm(\chapterlabel.\number\equanumber)}}#1}

\def\eqn{\eqno\eqname}

\appendix{A.}{}

In the standard Chern-Simons (CS) description of Anyons one introduces a CS
gauge field $\A$, whose field strength is proportional to the density:
$$
\vec\nabla\wedge\A\propto\rho
\eqn\
$$
In this formulation to have translational invariance one has to take care
of the boundary conditions in a proper way and to take a torus
\Kurt\ \Lidingping\ \Lechner\ \Iengo.
In this way one shows that the ground state of the full quantum solution of
the mean field theory corresponds to a constant density. Due to the
non-trivial topological properties of the torus one has to take into
account the topological components of the gauge potential (also called
``flat connections'' since $\vec\nabla\wedge\vec a\!=\!0$) defined by:
$$
a_x=\oint dx A_x \qquad a_y=\oint dy A_y
\eqn\
$$
here the two integrals are performed along the two non-trivial loops of the
torus.
\p
In this way the Hamiltonian can be written as follows
\Kurt\ \Lidingping\ \Lechner\ \Iengo:
$$
H=\int d^2x\int d^2a\,\biggl\{{1\over 2m}|\vec D\psi|^2 +
c\Bigl|\Bigl({k\over 4\pi}a_i+i\,\epsilon_{ij}{\pa\over\pa a_j}
\Bigr)\psi\Bigr|^2\biggr\}
\eqn\
$$
where the covariant derivative is:
$$
D_i=\pa_i-{i\pi\over 2k}\v\epsilon_{ij}x_j-i{a_i\over L}-i\AT_i
\eqn\
$$
here $L$ is the length of a side of the torus (we are supposing, for
simplicity that our torus is a square with identified edges), $\v$ is the
mean density, $\skew6\vec{\AT}$ is the fluctuation part of $\A$ such
that $\vec\nabla\wedge\skew6\vec{\AT}\!\propto\!\delta\rho$. $c$ is
a positive constant whose value can be arbitrary in what follows. In the
Anyon problem the value of $c$ is large, $c\!\to\!\infty$, and thus the
$a_i$ degrees of freedom remain in the ground state.
\p
In the case of the infinite plane we have to take the limit
$L\!\to\!\infty$ at constant density (we will call it {\it thermodynamical
limit}), so ${a_i\over L}\!\to\!0$ if $a_i$ is bounded (we will see below
that indeed $a_i$ is bounded) but in the Hamiltonian still survives a term
in $a_i$.
\p
Note that now the wavefunction $\psi$ is now a function of $\vec a$ beside
$\vec x$; \ie $\psi\!=\!\psi(\vec x, \vec a)$ so we define the density
$\rho(\vec x)$ to be:
$$
\rho(\vec x)=\int d^2a\,|\psi(\vec x, \vec a)|^2
\eqn\
$$
It is convenient to introduce a complex notation:
$$
A=A_M-i\AT \qquad \Aa=\Aa_M+i\AaT
\eqn\
$$
where
$$
A_M={i\over 2}{\pi\over k}\v\zz \qquad
\Aa_M=-{i\over 2}{\pi\over k}\v z
\eqn\
$$
are the ``mean field'' parts and
$$
\AT={i\over 2}(\AT_1-i\AT_2) \qquad \AaT=-{i\over 2}(\AT_1+i\AT_2)
\eqn\
$$
are the ``fluctuating'' parts; for the flat connections we define:
$$
a={1\over 2\pi}(ia_1+a_2) \qquad \aA={1\over 2\pi}(-ia_1+a_2)
\eqn\
$$
For the $CS$ magnetic field we have:
$$
B=2i(\pp A-\pa\Aa)=B_M+\BT=-{2\pi\over k}\rho
\eqn\
$$
where:
$$
B_M=-{2\pi\over k}\v \qquad \BT=(\pp \AT-\pa\AaT)=
{2\pi\over k}(\v-\rho)=-{2\pi\over k}\delta\rho.
\eqn\
$$
Let us now suppose that $\AT\!=\!0$ and study the mean field solution. The
Hamiltonian is:
$$
H=\int d^2z\int d^2a\,\biggl\{{2\over m}\bigl|\bigl(\pa+{\pi\over
2k}\v\zz\bigr)
\psi\bigr|^2+{\pi\over mk}\v |\psi|^2+
{c\over\pi^2}\Bigl|\Bigl({\pa\over\pa a}+
{\pi k\over 2}\aA\Bigr)\psi\Bigr|^2\biggr\}
\eqn\ste
$$
The state of minimal energy corresponds to:
$$
\psi=\psi_M=e^{-{\pi\v\over 2k}z\zz-{\pi k\over 2}a\aA}\, g(\zz,\aA)
\eqn\fano
$$
here $g$ is an arbitrary antiholomorphic function.
\p
{}From \fano\ we can see that $a$ is bounded.
\p
Let us look for the constant density solutions, which are known to
correspond to the ground state of the full quantum mechanical mean field
problem. Choosing $g(\zz,\aA)\!=\!e^{\pi v\zz\aA}\,v$ we get $\rho_M\!=\!\v$,
in fact
$$
\rho_M=\int d^2a\, |\psi_M|^2=\int d^2a \,e^{-\pi k|a-{v\over k}\zz|^2}\v=\v
\eqn\
$$
provided we normalize the measure $d^2a$ such that
$\int d^2a e^{-\pi k a\aA} =1$.
\p
For the Hamiltonian \ste\ we get:
$$
H_M={\pi\over mk}\v N
\eqn\
$$
this is the energy of $N$ particles in the lowest Landau level. We know that
in the mean field solution actually the $N$ particles fill exactly $k$
levels corresponding to the energy:
$$
E_M={1\over 2m}|B_M| Nk
\eqn\albe
$$
So, in order to reproduce the correct mean field energy, we must add:
$$
E'={1\over 2m}|B_M| N(k-1)=
{2\pi\over m}\Bigl(1-{1\over k}\Bigr)\int d^2a \int d^2z |\psi|^4
\eqn\
$$
So our correct starting Hamiltonian is:
$$
H=\int d^2x\int d^2a\,\biggl\{{1\over 2m}|\vec D\phi|^2+
{c\over\pi^2}\Bigl|\Bigl({\pa\over\pa a}+{\pi k\over 2}\aA\Bigr)\psi\Bigr|^2+
{2\pi\over m}\Bigl(1-{1\over k}\Bigr)|\psi|^4\biggr\}
\eqn\conte
$$
\medskip
\noindent
Notice that more in general a constant density is also obtained taking:
$$
\psi_M=e^{-{\pi\v\over 2k}z\zz-{\pi k\over 2}a\aA+
\pi v\zz\aA+i(pz+\bar p\zz)}\, v
\eqn\
$$
then the Hamiltonian \conte\ becomes:
$$
H_M=\int d^2z\biggl\{{2\over m}|ip|^2\v-{k\over 2m}B_M\v\biggr\}=
{1\over 2m}\bigl(p_x^2+p_y^2\bigr)N+{k\over 2m} |B_M| N
\eqn\
$$
so we have found, beside the standard mean field energy \albe\ a kinetic energy
equal to that of one particle times $N$. So our system moves like a
condensate where all particles have the same momentum. In other words it
represents a collective motion. If we compute the currents:
$$
J=\int d^2a\, {1\over 2mi}\Bigl[\psi^{\dag}D\psi-\psi(D\psi)^{\dag}
\Bigr]
\eqn\
$$
we get:
$$
J={\v\over m}p
\eqn\
$$
So we have found for the current exactly the charge density times the
velocity.
\bigskip
Now let us introduce the fluctuations taking $\AT\!\neq\!0$:
$$
H=\int d^2z\int d^2a\,\biggl\{{2\over m}\Bigl|\bigl(\pa+{\pi\over 2k}\v\zz-
\AT\bigr)\psi\Bigr|^2-{1\over 2m}B|\psi|^2+
{2\pi\over m}\Bigl(1-{1\over k}\Bigr)|\psi|^4+
$$
$$
+{c\over\pi^2}\Bigl|\Bigl({\pa\over\pa a}+
{\pi k\over 2}\aA\Bigr)\psi\Bigr|^2\biggr\}
\eqn\
$$
\goodbreak
Then if we take:
$$
\psi=e^{-{\pi\v\over 2k}z\zz-{\pi k\over 2}a\aA+
\pi v\zz\aA}\,\phi(z,\zz)
\eqn\
$$
for the density we get:
$$
\rho=\int d^2a\,|\psi|^2=\int d^2a\, e^{-\pi k|\aA-{v\over k}z|^2}
\,|\phi(z,\zz)|^2=|\phi(z,\zz)|^2
\eqn\
$$
and for the Hamiltonian:
$$
H=\II\biggl\{{1\over 2m}\bigl|\bigl(\vec\nabla-i\skew6\vec{\AT}\bigr)\phi
\bigr|^2+g|\phi|^4\biggr\}
\eqn\rto
$$
where in order to reproduce the correct ``mean field'' energy we have to
take $g\!=\!{\pi\over m}(1-{1\over k})$.
\bigskip
So we have found an Hamiltonian which in the ``mean field'' case gives the
correct answer and in the more general ``fluctuating'' case recovers the
covariant derivatives taking into account the density fluctuations.
\p
We take equation \rto\ as the basic effective Hamiltonian, describing our
quantum fluid, corresponding to the effective Lagrangian density \lagr.

\vfill
\eject

\newcount\equanumber	\equanumber=0
\def\chapterlabel{B}
\def\eqname#1{\relax\global\advance\equanumber by 1
  \xdef#1{{\rm(\chapterlabel.\number\equanumber)}}#1}

\def\eqn{\eqno\eqname}

\appendix{B.}{}

We consider the geometry of section 4.2, Fig. 2{\it b}), and a configuration
relevant for the penetration of the magnetic field in the $z$-direction.
\p
Assume that the system is homogeneous in the
$x$-direction. We suppose that $\phi$ is constant
up to a phase \ie $\phi\!=\!ve^{i\gamma}$, therefore we have no
electrostatic interaction and  put $\aCS\!=\!0$ (therefore this
configuration would be there also in a non chiral superconductor).
\p
In this configuration one considers the possibility that the phase is
responsible for the cancellation of $\aem$:
$$
\pa_y\gamma=e\aem
\eqn\relli
$$
We again suppose to have a fixed external flux so that equation
\uani\ still holds.
\p
We further impose periodic boundary conditions (ensuring the current
conservation):
$$
\gamma(0)=\gamma(L_y)=0({\rm mod}\, 2\pi)
\eqn\dista
$$
{}From \relli\ and \dista\ we get:
$$
eL_y\aem=2\pi p
\eqn\
$$
here $p\in {\bf Z}$ has in general a different value in different layers.
Since $d$ is the spacing between the layers the $z$ coordinate of the
$s$-th layer is $z_s\!=\!sd$. Now recalling equation \uani\ we get:
$$
eL_y\aem(z_s)=-eL_y{\varphi_0\over l_z}sd=2\pi p_s \qquad\qquad
\Rightarrow\qquad\qquad
p_s=-{e\over 2\pi}{d\over l_z}\varphi_0L_ys\in{\bf Z}
\eqn\
$$
but since the total flux is quantized $\varphi_0L_y\!=\!{2\pi\over e}n$ we
have:
$$
{p_s\over s}=-{d\over l_z}n
\eqn\quant
$$
since ${l_z\over d}$ is the number of layers ``penetrated'' by $B$ this
means that every layer bears an identical fraction of the total flux.
\p
The Hamiltonian of this system can be written as:
$$
H=\int dxdydz\biggl\{{1\over 2e^2\lambda^2}\Bigl|\pa_y\gamma-e\aem\Bigr|^2+
{1\over 2e^2\lambda_J^2}|\pa_z\gamma|^2+{1\over 2}(\pa_z\aem)^2\biggr\}
\eqn\
$$
the second term is the Josephson coupling between neighbouring layers
\Doniach. $\lambda_J$ is a constant of the dimension of a length whose
physical significance will be clear in the following, it is related to
the Josephson maximal current density $j_0$ by the equation:
$$
j_0={1\over ed\lambda_J^2}
\eqn\
$$
$\lambda$ is the standard penetration length \lx.
Since from \relli\ we have $\gamma\!=\!ey\aem$, we can rewrite the
hamiltonian as:
$$
H=L_x\int_0^{L_y}\int_0^{l_z}dz\biggl\{{1\over 2}(\pa_z\aem)^2
\biggl(1+{y^2\over \lambda_J^2}\biggr)\biggr\}=
{1\over 2}\varphi_0^2{L_xL_y\over l_z}
\biggl[1+{1\over 3}\biggl({L_y\over\lambda_J}\biggr)^2\biggr]
\eqn\cazzo
$$
which is minimal for $l_z$ macroscopic \ie for
$l_z\!\sim\! L_z\!\sim\! L_y\!\sim\! L_x$.
To see if this configuration is favoured one has to compare it with the
energy of the standard case in which:
$$
H^{{\rm standard}}=\varphi_0^2{L_xL_y\over\sqrt{3}\lambda}
\eqn\
$$
That is, for $l_z\!\sim\!L_y$, the configuration giving \cazzo\ is
favourite if:
$$
{1\over\sqrt{3}\lambda}>
{1\over 2l_z}\biggl[1+{1\over 3}\biggl({L_y\over\lambda_J}\biggr)^2\biggr]
\eqn\salva
$$
For all common high $T_c$ materials $\lambda_J\!\simeq\!10^5{\rm\AA}$
(corresponding to $j_0\!\simeq\!10^4 {\rm A/cm^2}$, see for example
\ref\procClem) therefore since ${\lambda_J\over\lambda}\!\simeq\! 120$, for
$l_z\!\sim\!L_y\mincir\!10^8{\rm\AA}$, this configuration could be
marginally competitive. When equation \salva\ is not satisfied then this
configuration would not be favourite.

\vfill
\eject

\centerline{{\tenbf Figure Captions}}
\bigskip

\noindent
1. Qualitative spectrum.
\p
2. Directions of the magnetic field with respect to the layers.

\vfill
\eject

\frenchspacing

\font\smc = cmcsc10

\centerline{{\tenbf References}}
\bigskip

\noindent
1. {\smc S.C. Zhang, T.H. Hansson, S. Kivelson},
{\it Phys. Rev. Lett.} {\bf 62} (1989), 82.
\vskip 0.3truecm

\noindent
2. {\smc D.H. Lee, S.C. Zhang},
{\it Phys. Rev. Lett.} {\bf 66} (1991), 1220.
\vskip 0.3truecm

\noindent
3. {\smc D.H. Lee, M.P.A. Fisher},
{\it Int. J. Mod. Phys.} {\bf B5} (1991), 2675.
\vskip 0.3truecm

\noindent
4. {\smc S.C. Zhang},
{\it Int. J. Mod. Phys.} {\bf B6} (1992), 25.
\vskip 0.3truecm

\noindent
5. {\smc D. Boyanovsky},
{\it Int. J. Mod. Phys.} {\bf A7} (1992), 5917.
\vskip 0.3truecm

\noindent
6. {\smc R. Iengo, K. Lechner},
{\it Nucl. Phys.} {\bf B346} (1990), 551.
\vskip 0.3truecm

\noindent
7. {\smc R. Iengo, K. Lechner, Dingping Li},
{\it Phys. Lett.} {\bf B269} (1991), 109.
\vskip 0.3truecm

\noindent
8. {\smc R. Iengo, K. Lechner},
{\it Phys. Rep.} {\bf 213} (1992), 179.
\vskip 0.3truecm

\noindent
9. {\smc R. Iengo, K. Lechner},
{\it Nucl. Phys.} {\bf B384} (1992), 541.
\vskip 0.3truecm

\noindent
10. {\smc J.R. Clem, M.W. Coffey},
{\it Phys. Rev.} {\bf B42} (1990), 6209.
\vskip 0.3truecm

\noindent
11. {\smc S.K. Paul, A. Khare},
{\it Phys. Lett.} {\bf B174} (1986), 420.
\vskip 0.3truecm

\noindent
12. {\smc J. Hong, Y. Kim, P.Y. Pac},
{\it Phys. Rev. Lett.} {\bf 64} (1990), 2230.
\vskip 0.3truecm

\noindent
13. {\smc R. Jackiw, E.J. Weinberg},
{\it Phys. Rev. Lett.} {\bf 64} (1990), 2234.
\vskip 0.3truecm

\noindent
14. {\smc E.B. Bogomol'nyi},
{\it Sov. J. Nucl. Phys.} {\bf 24} (1976), 449.
\vskip 0.3truecm

\noindent
15. {\smc R. Jackiw, S.Y. Pi},
{\it Phys. Rev. Lett.} {\bf 64} (1990), 2969.
\vskip 0.3truecm

\noindent
16. {\smc X.G. Wen, A.Zee},
{\it Phys. Rev. Lett.} {\bf 62} (1989), 2873.
\vskip 0.3truecm

\noindent
17. {\smc L.N. Bulaevskii},
{\it Int. J. Mod. Phys.} {\bf B4} (1990), 1849.
\vskip 0.3truecm

\noindent
18. {\smc J.R. Clem},
{\it Phys. Rev.} {\bf B43} (1991), 7837.
\vskip 0.3truecm

\noindent
19. {\smc J.M. Kosterlitz, D. Thouless},
{\it J. Phys.} {\bf C6} (1973), 1181.
\vskip 0.3truecm

\noindent
20. {\smc M.R. Beasley, J.E. Mooij, T.P. Orlando},
{\it Phys. Rev. Lett.} {\bf 42} (1979), 1165.
\vskip 0.3truecm

\noindent
21. {\smc J.R. Clem, M.W. Coffey, Z. Hao},
{\it Phys. Rev.} {\bf B44} (1991), 2732.
\vskip 0.3truecm

\noindent
22. {\smc L. Bulaevskii, J.R. Clem},
{\it Phys. Rev.} {\bf B44} (1991), 10234.
\vskip 0.3truecm

\noindent
23. {\smc J.R. Clem},
{\it Physica} {\bf C 162-164} (1989), 1197.
\vskip 0.3truecm

\noindent
24. {\smc W.E. Lawrence, S. Doniach},
{\it in} ``Proceedings of the  XII International  Conference on Low
Temperature Physics'' (E. Kanda, Ed.), p.361, Academic Press, Tokio, 1971.
\vskip 0.3truecm

\noindent
25. {\smc J.R. Clem},
{\it in} ``Physics and materials science of high temperature
superconductors'' (R. Kossowski, S. Methfessel, D. Wohlleben, Eds.)
NATO ASI Series E Vol. 181, p. 79, Kluwer Academic Publishers, Dordrecht, 1989.
\vskip 0.3truecm

\vfill
\eject
\bye